\documentclass[11pt,ams]{article}%
\usepackage{geometry}
\usepackage{dsfont}
\usepackage{amsmath}
\usepackage{amsfonts}
\usepackage{amssymb}
\usepackage{graphicx}%
\usepackage{subfig}
\usepackage{float}

\begin{document}

\title{\textbf{A few remarks on the zero modes of the Faddeev-Popov operator in the Landau and maximal Abelian gauges}}

\author{
\textbf{M.~S.~Guimaraes}$^{a}$\thanks{msguimaraes@uerj.br}\,\,,
\textbf{S.~P.~Sorella}$^{a}$\thanks{sorella@uerj.br}\ \thanks{Work supported by
FAPERJ, Funda{\c{c}}{\~{a}}o de Amparo {\`{a}} Pesquisa do Estado do Rio de
Janeiro, under the program \textit{Cientista do Nosso Estado}, E-26/101.578/2010.}\,\,,\,\,
\\[2mm]
\textit{{\small {$^{a}$ UERJ $-$ Universidade do Estado do Rio de
Janeiro}}}\\\textit{{\small {Instituto de F\'{\i}sica $-$
Departamento de F\'{\i}sica Te\'{o}rica}}}\\\textit{{\small {Rua
S{\~a}o Francisco Xavier 524, 20550-013 Maracan{\~a}, Rio de
Janeiro, RJ, Brasil.}}}$$}
\maketitle

\date{}

\begin{abstract}
\noindent The construction outlined by Henyey  \cite{Henyey:1978qd}  is employed to provide examples of normalizable zero modes of the Faddeev-Popov operator in the  Landau and maximal Abelian gauges in $SU(2)$ Euclidean Yang-Mills theories in $d=3$ dimensions. The corresponding gauge configurations have all finite norm $||A||^2 < \infty$. In particular, in the case of the Landau gauge, the explicit construction of an infinite class of normalizable zero modes with finite norm  $||A||^2 $ is provided.

\end{abstract}

\baselineskip=13pt

\newpage

\section{Introduction}
The characterization of the zero modes of the Faddeev-Popov operator in various gauges is relevant for the issue of the
Gribov copies \cite{Gribov:1977wm}, which affect in a non-perturbative way the  quantization of Yang-Mills theories. Zero modes of the Faddeev-Popov
operator give rise to gauge field configurations which are in fact Gribov copies. \\\\The issue of the zero modes of the Faddeev-Popov operator is usually
addressed by focusing on a specific gauge configuration as, for example, an instanton configuration \cite{Bruckmann:2000xd}. A different strategy to face the
explicit construction of Gribov copies in the Coulomb gauge has been outlined by Henyey \cite{Henyey:1978qd}, see also \cite{Sobreiro:2005ec}  for a pedagogical
review of the method  and \cite{Canfora:2010zf} for a recent application to Abelian gauge theories in curved backgrounds. Within Henyey's construction, both the zero mode and the corresponding gauge field configuration are determined in a self-consistent way. More precisely, the differential equation for the zero mode is employed as
an algebraic equation allowing us to express the gauge field  in terms of the function which parametrizes the zero mode itself. In this way, Henyey has been able
to provide examples of gauge copies with the desired asymptotic behavior for the resulting gauge field configuration. \\\\The aim of this work is twofold. First,  we revise Henyey's construction in the case of $SU(2)$ Landau gauge in $d=3$ dimensions. Here, we shall be able to provide the construction of an infinite class of normalizable zero modes whose corresponding gauge configurations have all finite norm $||A||^2 = Tr \int d^3x |\overrightarrow{A}|^2 < \infty$. Second, we point out that Henyey's strategy can be employed to obtain examples of zero modes in other gauges as, for example, the maximal Abelian gauge, which proves to be useful in order to investigate the dual superconductivity mechanism for confinement. In this case, despite the complexity of the Faddeev-Popov operator, Henyey's set up turns out to be helpful in order to obtain in a simple way an example of a normalizable zero mode whose corresponding gauge field has finite norm  $||A||^2  < \infty$. \\\\Let us remind here that the condition of finite norm $||A||^2$ plays an important role in the understanding of the Gribov issue which has been achieved so far. It turns out in fact that the requirement that the functional $f_A[U] = Tr \int d^3x |\overrightarrow{A}^U|^2$, with $A^U_i = {U}^{\dagger }\partial _{i}{U}+U^{\dagger }A_{i}{U}$, is well defined and differentiable along the gauge orbit of $A_i$ is at the basis of well established results concerning the properties of the Gribov and fundamental modular regions in the Landau gauge, see \cite{semenov,Dell'Antonio:1991xt,vanBaal:1991zw}. The Gribov region corresponds in fact to the set of all relative minima of the functional  $f_A[U]$ in the space of the gauge orbits, while the fundamental modular region is defined as the set of all absolute minima of  $f_A[U]$. In principle, the fundamental modular region is known to be free of Gribov copies. As such, the restriction of the domain of integration in the Feynman path integral to that region would  provide the correct way of quantizing Yang-Mills theories. Though, till now, a practical way of implementing such a restriction is still lacking. Nevertheless, it turns out that the restriction of the domain of integration to the Gribov region can be implemented within a local and renormalizable framework, although this region is still plagued by the presence of residual copies. This has resulted in the so called Gribov-Zwanziger action \cite{Zwanziger:1988jt,Zwanziger:1989mf}  and its more recent refined version \cite{Dudal:2007cw,Dudal:2008sp}. We underline that the functional $f_A[U]$ can also be introduced in the maximal Abelian gauge, where a few properties of the corresponding Gribov region have been established \cite{Capri:2008vk}. \\\\The present work is organized as follows. In Sect.2, Henyey's construction is employed in the case of the Landau gauge, where an infinite class of normalizable zero modes exhibiting finite norm $||A||^2$ is presented. Sect.3 is devoted to discuss Henyey's set up in the case of the maximal Abelian gauge.

\section{Henyey's construction of the zero modes in the Landau gauge}

Let us start by recalling the relationship between zero modes of the Faddeev-Popov operator and Gribov copies. Let $A_i$  denotes a $SU(2)$ Lie-algebra valued field in $d=3$ Euclidean space, $i=1,2,3$, obeying the Landau gauge condition  $\partial_{i}A_{i}=0$. Under a gauge transformation,  the  field $A_i$ transforms as
\begin{equation}
\widetilde{A}_{i}={U}^{\dagger }\partial _{i}{U}+U^{\dagger }A_{i}{U}\;.  \label{eq1a}
\end{equation}
For an infinitesimal gauge transformation, $U=1+\alpha$, eq.\eqref{eq1a} reduces to
\begin{equation}
\widetilde{A}_{i}=A_{i}+\left( \partial _{i}\alpha
+\left[ A_{i},\alpha \right] \right) \;. \label{eq1b}
\end{equation}
The gauge transformed field $\widetilde{A}_{i}$ is  a Gribov copy of $A_i$ if it fulfills the Landau gauge condition $\partial_{i}\widetilde{A}_{i}=0$. We see thus, from eq.(\ref{eq1b}), that this is the case if there exists a  $SU(2)$ Lie-algebra valued function $\alpha = \omega$ such that:
\begin{equation}
\nabla^{2}\omega +\left[ A_{i},\partial _{i}\omega
\right] =0\;.  \label{eq1}
\end{equation}
The differential operator ${\cal M} = (\nabla^2\cdot + [A_i, \partial_i\;\cdot ])$ appearing in eq.\eqref{eq1} is nothing but the Faddeev-Popov operator of the Landau gauge, and $\omega$ is a zero mode of ${\cal M}$. \\\\In order to construct normalizable solutions of eq.\eqref{eq1}, we shall follow the strategy outlined by Henyey \cite{Henyey:1978qd} and consider the following gauge configuration:
\begin{equation}
\overrightarrow{A}=i\overrightarrow{a}\sigma _{3}\;,\qquad \;\;\;\overrightarrow{\nabla}\cdot \overrightarrow{A}=0\;,  \label{eq2}
\end{equation}
that is, we are considering a particular transverse gauge field configuration along the abelian direction of the gauge group $SU(2)$ defined by the Pauli matrix $\sigma_3$. Adopting spherical coordinates $(r,\theta,\varphi)$, the Landau gauge condition is fulfilled by setting
\begin{equation}
\overrightarrow{a}=a(r,\theta )\hat{e}_{\varphi
}\;.\;\;\; \label{eq3}
\end{equation}
We next consider the following ansatz \cite{Henyey:1978qd} for  the zero mode $\omega$
\begin{equation}
\omega (r,\theta, \varphi)= r b(r) \sin\theta\left( \sigma _{1}\cos \varphi +\sigma
_{2}\sin \varphi \right) \;,  \label{eq4}
\end{equation}
where the radial function $b(r)$ is left free. Using eqs.\eqref{eq3}, \eqref{eq4}, the differential equation \eqref{eq1} becomes
\begin{equation}
(4b^{\prime }+rb^{\prime \prime })\sin \theta +ab=0\;.
\label{eq5}
\end{equation}
It is thus apparent that this equation can be seen as an algebraic equation enabling us to express the field configuration $a$ in eq.\eqref{eq3} in terms of the function $b$ parametrizing the zero mode $\omega$, namely
\begin{equation}
a(r,\theta)=-\frac{r}{b}\sin \theta \left( b^{\prime \prime }+\frac{4b^{\prime }}{r}%
\right) \;.  \label{eq6}
\end{equation}
We suffice now to find a function $b(r)$ providing a normalizable zero mode $\omega$ and a gauge field $A_i$ with finite norm $||A||^2<\infty$.\\\\To that purpose we choose  the following form for the function $b(r)$:
\begin{equation}
b(r)=\frac{k}{r^3+\alpha r^\gamma+\beta}\;,  \label{eq7}
\end{equation}
where $k$, $\alpha$, $\beta$ and $\gamma$ are real constants, and we initially consider $0<\gamma<3$. Expression  \eqref{eq7} can be seen as a generalization of the original Henyey's proposal \cite{Henyey:1978qd}.  Note that this function is regular at the origin $r=0$ if $\beta \neq 0$ and thus, by eq.(\ref{eq4}), the zero mode $\omega$ is also regular. The function $a(r,\theta)$ defining the gauge field configuration can be easily evaluated by means of eq.(\ref{eq6}), and reads
\begin{equation}
a(r,\theta)= \frac{\sin\theta}{(r^3+\alpha r^\gamma+\beta)^2}\left[ \alpha(\gamma^2 -9\gamma+18)r^{\gamma+2} - \alpha^2 (\gamma^2-5\gamma)r^{2\gamma-1}+\alpha\beta\gamma(\gamma+3)r^{\gamma-1}+18\beta r^2\right]\;.  \label{eq8}
\end{equation}
We see thus that the gauge field is also regular at the origin provided $\gamma \geq 1$.\\\\Let us check that the zero mode $\omega$ is normalizable, {\it i.e.}
\begin{equation}
||\omega||^2 = Tr \int d^{3}x\;  \omega^{\dagger }\omega\; < \infty \;.
\label{eq9}
\end{equation}
In the present case we have $Tr \omega^{\dagger }\omega = 2r^2b^2\sin^2\theta$, so that
\begin{equation}
Tr \int d^{3}x\;  \omega^{\dagger }\omega\; = 4\pi k^2\int_0^{\pi}d\theta\sin^3\theta\int_0^{\infty}dr\frac{r^4}{(r^3+\alpha r^\gamma+\beta)^2} < \infty \;.
\label{eq10}
\end{equation}
Since the radial integrand goes as $1/r^2$ for $r\rightarrow \infty$ (for $1 \leq \gamma<3$), expression \eqref{eq10} shows that  we have in fact a normalizable zero mode.\\\\Likewise, the gauge field norm $||A||^2$ is also required to be finite, {\it \i.e.}
\begin{equation}
||\overrightarrow{A}||^2 =Tr  \int d^{3}x\;  |\overrightarrow{A}|^2\; < \infty \;.
\label{eq11}
\end{equation}
From eq.(\ref{eq8}) we have:
\begin{equation}
||\overrightarrow{A}||^2 = 4\pi\int_0^{\pi}d\theta\sin^3\theta\int_0^{\infty}dr\frac{\left[ \alpha(\gamma^2 -9\gamma+18)r^{\gamma+3} - \alpha^2 (\gamma^2-5\gamma)r^{2\gamma}+\alpha\beta\gamma(\gamma+3)r^{\gamma}+18\beta r^3\right]^2}{(r^3+\alpha r^\gamma+\beta)^4}\;.
\label{eq12}
\end{equation}
Note that the requirement $\gamma<3$ establishes the first term in the numerator as the dominant one for large $r$. Thus, for $r\rightarrow\infty$, the radial integrand behaves as $r^{2\gamma-6}$ and the integral will be finite if
\begin{equation}
2\gamma-6 < -1 \Rightarrow \gamma < \frac52 \;.
\label{eq13}
\end{equation}
Therefore, choosing the radial function $b(r)$ as
\begin{equation}
b(r)=\frac{k}{r^3+\alpha r^\gamma+\beta}\;, \qquad        1\leq \gamma < \frac{5}{2} \;,   \label{br}
\end{equation}
will lead to an infinite class of normalizable zero modes, corresponding to the infinite set of values of the exponent
$\gamma$. The related gauge field configurations have all finite norm $||A||^2<\infty$.

\section{Examples of zero modes in the maximal Abelian gauge}

The aim of this section is that of showing that Henyey's strategy can be successfully employed also in the  case of the maximal Abelian gauge. Here, the gauge field is split into  diagonal  and off-diagonal components. In the case of $SU(2)$, the diagonal component corresponds to the diagonal generator of the Cartan subgroup, {\it i.e.} $\sigma^3$, and is identified with $A^3_i$, while the off-diagonal components $A_i^{a}$, with  $a=1,2$, correspond to the remaining off-diagonal generators, {\it i.e.} $\sigma^a$.   In this case, different gauge conditions are imposed on  the diagonal and off-diagonal components, namely  \begin{equation}
D_i^{ab}A^b_i = 0\;,\;\;\;\partial_i A_i^{3}=0\; ,
\label{eq2mag}
\end{equation}
where
\begin{equation}
D_i^{ab} = \delta^{ab}\partial_i - \varepsilon^{ab}A^3_i \;,
\label{eq3mag}
\end{equation}
is the covariant derivative with respect to the diagonal component $A^3_i$.
For the Faddeev-Popov operator we have now  \cite{Capri:2008vk}
\begin{equation}
{\cal M}^{ab} = - \left( D_i^{ac}D_i^{cb}  + \varepsilon^{ac}\varepsilon^{bd}A^c_iA^d_i  \right)\;, \label{magfp}
\end{equation}
so that the differential equation for a zero mode reads
\begin{eqnarray}
D_i^{ac}D_i^{cb} \alpha^b + \varepsilon^{ac}\varepsilon^{bd}A^c_iA^d_i\alpha^b = 0\:.
\label{eq4mag}
\end{eqnarray}
In order to construct examples of zero modes we shall consider the following gauge field configuration
\begin{equation}
A^3_i =0 \;, \qquad A^c_i = \delta^{1c} a_i \;,
\label{eq6mag}
\end{equation}
The maximal Abelian gauge fixing conditions, eqs.\eqref{eq2mag},  reduce to
\begin{equation}
{\overrightarrow\nabla}\cdot \overrightarrow{a}=0\;.
\label{eq7mag}
\end{equation}
Proceeding as in the previous section, we adopt spherical coordinates $(r,\theta,\varphi)$, so that condition \eqref{eq7mag}
is fulfilled  by setting
\begin{equation}
\overrightarrow{a}=a(r,\theta )\hat{e}_{\varphi
}\;.\;\;\; \label{eq8mag}
\end{equation}
The differential equation for the zero mode, eq.\eqref{eq4mag}, becomes thus
\begin{eqnarray}
\nabla^2 \alpha^{(1)} = 0\:,\nonumber\\
\nabla^2 \alpha^{(2)} + a^2 \alpha^{(2)} =0 \;.
\label{eq9mag}
\end{eqnarray}
Setting $\alpha^{(1)}=0$ and $\alpha^{(2)}=\omega$, one immediately realizes that  the gauge field configuration $a$  can be expressed in terms of $\omega$, according to
\begin{eqnarray}
a^2 = - \frac{1}{\omega}\nabla^2 \omega\;.
\label{eq10mag}
\end{eqnarray}
Therefore, we are in a situation very similar to that of the Landau gauge, and Henyey's set up can be easily adapted to the present case.
Adopting in fact the following ansatz for $\omega$
\begin{eqnarray}
\omega (r,\theta, \varphi)= r b(r) \sin\theta\cos \varphi\;,
\label{eq11mag}
\end{eqnarray}
it follows that the gauge configuration is fully determined by the radial function $b(r)$, namely
\begin{equation}
a^2 =-  \frac{1}{rb}\left(rb^{\prime \prime }+4b^{\prime}\right)\;.
\label{eq12mag}
\end{equation}
It suffices thus to find a suitable function $b(r)$ which gives rise to a normalizable zero mode while providing a finite norm for the gauge configuration. It is worth to notice here the interesting feature that equation \eqref{eq12mag} directly provides the norm of the gauge field, namely
\begin{equation}
|| A ||^2 = 2 \int d^3x\; a^2 \;. \label{norm}
\end{equation}
 In the present case, it turns out that a simple choice for the  function $b(r)$ is provided by
 \begin{equation}
 b(r) =  \frac{k}{r^m+\beta}\;, \label{bf}
 \end{equation}
where $(k,\beta)$ are non-vanishing constants and the exponent  $m>\frac{5}{2}$ in order to ensure that the zero mode $\omega$  has finite norm, {\it i.e.}
\begin{equation}
	\int d^3x\;  \omega^2   \propto  \int_0^\infty dr\;\frac{r^4}{(r^m+\beta)^2} < \infty  \qquad  {\it for \;\; m>\frac{5}{2} }   \;. \label{normzm}
\end{equation}
From expression \eqref{bf} it turns out that
\begin{equation}
a^2 = \frac{1}{(r^m+\beta)^2} \left(  (3m-m^2) r^{2m-2} + \beta (m^2+3m) r^{m-2} \right)  \;, \label{a2}
\end{equation}
from which it follows that the gauge configuration $a$ is regular at the origin $r=0$, while
\begin{equation}
a^2_{r\rightarrow \infty} \sim \frac{1}{r^2} \;, \label{mnz}
\end{equation}
unless $m=3$, in which case
\begin{equation}
a^2 =  \frac{18 \beta  r}{(r^3+\beta)^2}   \;, \qquad \; {\it and } \;\qquad
a^2_{r\rightarrow \infty} \sim \frac{1}{r^5} \;,\label{a2f}
\end{equation}
giving rise to a finite norm $||A||^2$. \\\\As such,  expression
\begin{equation}
b(r) = \frac{k}{r^3+\beta}\;, \label{bf3}
\end{equation}
gives rise to a normalizable zero mode whose corresponding gauge field $a$, eq.\eqref{a2f}, has finite norm, showing thus that Henyey's set up can be successfully employed  in the maximal Abelian gauge.
\subsection{Relaxing the finite norm requirement}
Let us conclude this section with a remark on the expression \eqref{a2}. For a generic value of the exponent $\frac{5}{2}<m<3$ (this requirement, $\frac{5}{2}<m<3$, is needed in order to ensure that $a^2$ in expression \eqref{a2} does not become negative for large $r$), the resulting gauge configuration $a$ does not provide a finite norm $||A||^2$, due to the asymptotic behavior \eqref{mnz}. However, it is  interesting to observe that such a configuration will give rise to a finite Yang-Mills action. In fact, a simple calculation shows that
\begin{equation}
S_{YM} = \frac{1}{4g^2} \int d^3x \left(  F^3_{ij} F^3_{ij} + F^a_{ij} F^a_{ij}  \right) = \frac{2\pi}{g^2} \int_0^\infty dr\;r^2 \frac{\partial a}{\partial r}  \frac{\partial a}{\partial r}   \;, \label{fact}
\end{equation}
which is finite since, according to  expression \eqref{mnz}, the gauge configuration $a$ behaves as $1/r$ for $r\rightarrow \infty$. In summary, although not exhibiting a finite norm $||A||^2$ for a generic value of the exponent $\frac{5}{2}<m<3$, expressions \eqref{bf} and \eqref{a2} provide an example of an infinite class (labelled by the exponent $m$) of normalizable zero modes with finite Yang-Mills action. An analogous situation is easily found also  in the case of the Landau gauge.
\section{Conclusion}
In this work, Henyey's set up \cite{Henyey:1978qd} has been employed to construct examples of normalizable zero modes of the Faddeev-Popov operators
corresponding to the Landau and maximal Abelian gauges in $SU(2)$ Yang-Mills theories in $d=3$ dimensions. In particular, in the Landau gauge, an
infinite class of normalizable zero modes whose corresponding gauge configurations have finite norm $||A||^2$ has been provided.  \\\\Let us conclude with a few
general remarks on the issue of the Gribov copies:
\begin{itemize}

\item needless to say, the examples of the zero modes and related gauge field configurations which we have been able to construct in a rather easy way give a simple confirmation of the pivotal relevance and, at the same time, of the complexity of the issue of the Gribov copies for a correct quantization of Yang-Mills theories.

\item although a full resolution of the Gribov issue is still lacking, we would have a tendency to believe that the {\it thermodynamical/statistical} treatment of this issue, as started by Gribov  \cite{Gribov:1977wm} and systematized by Zwanziger \cite{Zwanziger:1988jt,Zwanziger:1989mf}, seems very suitable. In a rough pictorial way, one could figure out that the infinite dimensional space of the Gribov copies could be associated to a kind of {\it thermodynamical reservoir} to which Yang-Mills theories are intrinsically connected. In this sense, the emergence of the dynamical massive Gribov parameter $\gamma^2$   \cite{Gribov:1977wm,Zwanziger:1988jt,Zwanziger:1989mf}  looks quite natural. To some extent, this parameter would encode the information of the existence of the infinite dimensional reservoir of Gribov copies, as much as the temperature $T$ carries the information of the presence of a thermodynamical bath. It is remarkable that the Gribov-Zwanziger treatment of the Gribov issue, amounting to cut off the domain of integration in the functional integral to the Gribov region, has resulted in a local and renormalizable action   \cite{Zwanziger:1988jt,Zwanziger:1989mf}. More recently, a refined version of the Gribov-Zwanziger action has been worked out \cite{Dudal:2007cw,Dudal:2008sp}. It is worth mentioning that, till now, the predictions of the refined action on the infrared behavior of the gluon and ghost two point correlation functions are in very good agreement with the most recent lattice data \cite{Cucchieri:2007rg,Cucchieri:2010xr,Bogolubsky:2009dc,Dudal:2010tf}.   Moreover, the refined action has provided rather good estimates for the lightest glueball states \cite{Dudal:2010cd}.

\item Finally, it is worth to spend a few  words  on the possible questioning which our present results might rise with respect to the frameworks which are available so far in order to face the issue of the Gribov copies. Here, we shall focus on the Landau gauge, due to the relatively large amount of results which have been already established on the issue of the Gribov copies  \cite{semenov,Dell'Antonio:1991xt,vanBaal:1991zw}. Our first comment is devoted to the fundamental modular region, which is defined as the set of all absolute minima of the functional  $f_A[U] = Tr \int d^3x |\overrightarrow{A}^U|^2$ along the gauge orbits. It is known that the fundamental modular region is free from Gribov copies   \cite{semenov,Dell'Antonio:1991xt,vanBaal:1991zw}. Moreover, the very definition of this region relies on the existence of the norm $||A||^2 $. On the other hand, we have seen that it is rather easy to construct copies which do not exhibit finite norm $||A||^2 $. Rather, the corresponding field configurations have finite action, which is believed to be a more physical requirement than that of finite norm. Till now, we are unaware of a physically motivated argumentation in order to demand that the relevant asymptotic behavior of the gauge fields at infinity should be that required by the finiteness of the norm $||A||^2 $. Moreover, we should also remind that, at present, a practical implementation of the restriction of the domain of integration of the functional integral to the fundamental modular region has not yet been achieved. Here, it seems safe to state that much work is needed to reach a satisfactory understanding of that region and of the physical consequences of  its implementation in the functional integral. \\\\Our second remark concerns the so called Gribov-Zwanziger framework \cite{Gribov:1977wm,Zwanziger:1988jt,Zwanziger:1989mf}, which allows one to effectively restrict, within a local and renormalizable field theory framework,  the functional integral to the Gribov region $\Omega$
\begin{equation}
\Omega = \{ \;A^a_\mu\;;\;\; \partial_\mu A^a_\mu=0 \;, \;\;- \partial_\mu D^{ab}_\mu >0 \; \}  \;. \label{Om}
\end{equation}
The boundary, $\partial \Omega$, of the region $\Omega$ is the first Gribov horizon. As it stands, expression \eqref{Om} has to be further characterized by specifying the asymptotic behavior at infinity of the gauge field configurations. This is usually done by stating that the region $\Omega$ corresponds to the set of all realtive minima of the functional $f_A[U] = Tr \int d^3x |\overrightarrow{A}^U|^2$ along the gauge orbits, a definition which establishes a relationship between  $\Omega$ and the  requirement of finite norm  $||A||^2 $. Once defined in that way, it is possible to prove the important result that every gauge orbit crosses $\Omega$ at least once   \cite{semenov,Dell'Antonio:1991xt,vanBaal:1991zw}, a result which provides a well defined support for restricting the functional integral to the region $\Omega$. Though, in principle, one could argue that, contrary to the fundamental modular region, the introduction of the region $\Omega$ could be done without any reference to the requirement of finite norm  $||A||^2 < \infty$.  It is apparent in fact from eq.\eqref{Om} that the region $\Omega$ could be introduced without any reference to the norm  $||A||^2$, which could be eventually replaced by the criterion of finite action. This will endow the Gribov region $\Omega$ with a different asymptotic behavior for the fields. However, as a consequence, the statement that every gauge orbit crosses the Gribov region should be completely re-investigated {\it ab initio}. Still, we might argue that the restriction to the region $\Omega$, endowed now with the criterion of finite action, would enable us to get rid at least of all copies related to zero modes, as it follows from the positivity of the Faddeev-Popov operator. As such, the so called Gribov-Zwanziger action would still retain a relevant meaning, due to the absence of zero modes. Though, important properties of the region $\Omega$, derived by making use of the norm $||A||^2$, should be completely re-examined, a task which does not look that easy. Evidently, we cannot be exhaustive here. Furthermore, we hope that the reader might get convinced of the crucial role that the issue of the Gribov copies represents for Yang-Mills theories. At the same time, we hope to have given a rough idea of how far we are from a satisfactory resolution of this issue.

\end{itemize}

\section*{Acknowledgments}
The Conselho Nacional de Desenvolvimento Cient\'{\i}fico e
Tecnol\'{o}gico (CNPq-Brazil), the Faperj, Funda{\c{c}}{\~{a}}o de
Amparo {\`{a}} Pesquisa do Estado do Rio de Janeiro, the Latin
American Center for Physics (CLAF), the SR2-UERJ,  the
Coordena{\c{c}}{\~{a}}o de Aperfei{\c{c}}oamento de Pessoal de
N{\'{\i}}vel Superior (CAPES)  are gratefully acknowledged.

\end{document}